# Towards humane digitization: a wellbeing-driven process of personas creation

**Irawan Nurhas**
Hochschule Ruhr West,
Bottrop, Germany.
University of Jyväskylä, Finland.
irawan.nurhas@hs-ruhrwest.de

**Jan M. Pawlowski**
Positive Computing Institute of
Hochschule Ruhr West
Bottrop, Germany
Jan.pawlowski@hs-ruhrwest.de

**Stefan Geisler**
Positive Computing Institute of
Hochschule Ruhr West
Bottrop, Germany
stefan.geisler@hs-ruhrwest.de

**ABSTRACT**
Digital transformation is a process of digitizing the working and living environment in which people are at the center of digitization. In this paper, we present a personas-based guideline for system developers on how the humanization of digital transformation integrates into the design process. The proposed guideline uses the positive personas from the beginning as a basis for the transformation of the working environment into the digital form. We used the literature research as a preliminary study for the process of wellbeing-driven digital transformation design, consisting of questions for structuring the required information in the positive personas as well as a potential method that could be integrated into the wellbeing-based design process.

**ACM Classification Keywords**
• Human-centered computing ~ User studies

**INTRODUCTION**
The rapid development of a technology that is accompanied by the benefits for the economy leads to the implementation of digital transformation. Digitization of work and living environments mainly focuses on the use of digital technology and accidentally forgets about the users of the technology itself. Therefore, it is important to guide the digital transformation process by focusing on people start from the beginning, digitizes not only the work but also humanize the digital design. In this study, we focused on well-being as a factor that humanizes the design of a system. The integration of well-being factors is implemented by adapting the personas that are commonly used in the design process.

Human-centered design expects to meet the requirements by taking usage behaviors in the design process. Cooper et al. recommend the use of personas as a tool to model usage behaviors [6]. In the goal-directed design, Cooper et al. use personas as a modeling tool for user behaviors, goals, and requirements [6]. With the upcoming trend towards positive design [9] and to promote human well-being in the development of information and communication systems [30], it is essential to integrate factors that affect the well-being into the system design [21]. Because the system design, including the user experience design, has a high impact on the economic value of business functions [24], it becomes more important to facilitate human well-being factors into the design. Focusing on human wellbeing in the design can result in an increased value proposition [33] and should lead to higher profitable returns [34] as well as a higher probability of a long-term relationship with the customer [11]. However, to our knowledge, there is still a lack of integration of well-being into personas as the bridge between research and design and as the communication tool between cross-functional departments in the development phase of system design. Furthermore, Desmet and Pohlmeyer proposed to develop an appropriate hands-on method for a designer to integrate well-being in the early stage of the design process [9]. We used a structured literature review [37] to re-conceptualize aspect in the personas that can be used to enhance the integration of wellbeing determinants into the system design process.

Based on the implementation gap as well as the potential of well-being and personas, this study aims to answer the research question about how the humanization of digital transformation works in the design process. In this study, we provide a wellbeing-driven design process based on positive personas [29] and propose a positive design intervention in the creation process of personas.

**STATE OF THE ART**
We start our work by presenting studies on the definition of humanizing the design and the product. Secondly, we present literature about the fundamental principles of well-being in the design process as well as the use of positive computing to design an information and communication system aiming to increase human well-being. Then, the use, creation process, and integration of personas in a different context.

### Humanizing Design Through Wellbeing-driven Approach

The first question that arises about the concept of humanization in design is the meaning of humanization itself, then what has been implemented and what potential has not been maximized in the application of humanization in design. The humanization of digital technology can be understood as a process of developing a technology-oriented product or service for people, by taking into account the qualities or values of the human aspect for a near-realistic use and environment. [17]. For this reason, it is essential to consider the values that guide the humanization process. The human-centric design has already focused on humanizing values in the design aspects such as usability, functionality, reliability and emotional aspects of design [12]. However, humanization concerning design [8], involves human factors such as empathy, multisensory optimization and attention to holistic thinking in design aspects [8], and all these aspects relate to human well-being, which goes hand in hand with a broader aspect.

There are still unused factors of well-being that can lead to focusing on more than the "wow" aspect of design but also on the "humanizing" dimension of design, such as meaningful goals, positive emotions, achievement and gratitude [5]. Danko explains the importance of a new method for the design, which focuses on the emotional connection between design and person [8].

### Well-being Model and the Principles in the System Design

Design for well-being attempts to deliver pleasurable and meaningful products [9]. Regarding well-being, Diener et al. defined well-being as momentary pleasure or fulfillment of life satisfaction [10]. Dodge et al. described well-being as a balanced position between a person's resources and challenges [13]. In short, we summarized that well-being is any condition that causes positive emotional reactions in the short- or long-term for an individual reason. Desmet and Pohlmeyer described a design for well-being as an innovative approach to new design opportunities that enables people to thrive and create a lasting effect on people's lives, and they also provide a positive design framework as a foundation for well-being in design [9]. Miller and Kalviainen proposed three ways to promote well-being in design: 1) improving awareness of prediction and control; 2) encourage social interaction; and 3) promote mindfulness, physical involvement, and enjoyment [28]. Design for well-being can help designers, not only to design innovative products but also to help them collaborate with marketers in the development phase of product design [33].

In the design of the technological environment, a positive computing approach based on well-being was introduced as a concept that can change the design approach of organizations, processes, and systems by focusing more on actors/stakeholders rather than the technology itself [30]. Positive computing provides well-being elements that affect the design of technology; ten essential elements of well-being have been identified [5] and could be utilized as a mediator of well-being (well-being determinant) in technological usage. Furthermore, a positive computing framework [5] and design principles for well-being [9, 28, 33] could be used as guidelines to integrate element well-being into a design tool.

Positive design and positive computing are derived from positive psychology, aimed to enhance human flourishing [5, 9, 23]. Lyubomirsky & Layous provide a positive activity model (PAM) that describes a simple positive action to improve human well-being by focusing attention on activity features and personal features [23]. The PAM concludes that well-being can be increased by practicing small activities to fulfill basic psychological needs [32]. However, demographic variables may affect the results of PAM to enhance well-being. This model can be used for our study since we mainly focus on a person and their behavioral attributes.

### The Personas in Action

Since the introduction of personas into Human-Computer Interaction (HCI) by Cooper et al. [6], Miaskiewicz and Kozar summarized the benefits of personas. For instance: the use of personas as a communication tool to discuss design decisions and as a tool to measure the design effectiveness, personas also can build commitment and consensus, by providing a simple way to communicate user behavior [27]. It has been reported that by adopting personas, the development team was gaining benefits [16, 27] and they used personas both for design and as a communication tool [25].

The creation process and attributes of personas have evolved, depending on the context of their use. Popular behavioral attributes of personas (activities, attitudes, aptitudes, motivations, and skills) have been used to model the user and the creation process by Cooper et al. as a fundamental process of personas creation [6]. Furthermore, in the field of HCI, the human-artifact model (HAM) was used to describe the relationship in technology-usage activities between a usage artifact and the human aspect [2]. Since the design process of HCI is a goal-oriented project, it can be easy to understand the importance of the project goal to drive the system development. The collective goals of the development team in the creation of the personas can improve cross-functional collaboration and integrating collective goals into the creation process [36] of personas is the critical aspect of this study. Therefore, in the following subsection, we present well-being as a collective goal of the development team to create personas.

### RE-CONCEPTUALIZING POSITIVE PERSONAS

In this section, we reconstructed the concept of positive personas. First, we present the method and the two main results: the guiding questions for the wellbeing-driven personas creation and the interventions of the positive personas in the development process of personas. In this study, we performed a structured literature review [37] to explore and build the intervention process of wellbeing-driven design based on positive personas.

We applied the period from 2000 to 2016 on 20 January 2017 and only selected articles written in English. We found related articles with keywords, such as "personas," "persona," and "well-being model," from the Association for Computing Machinery (ACM) and Springer Link. We also utilized the ISI Web of Knowledge and Google Scholar to find a specific topic related to well-being in design, for instance: "design well-being," and "positive computing." We read each abstract carefully to filter inappropriate articles. Then, we analyzed 89 articles for our study purpose; the complete paper selection process can be seen in Figure 1.

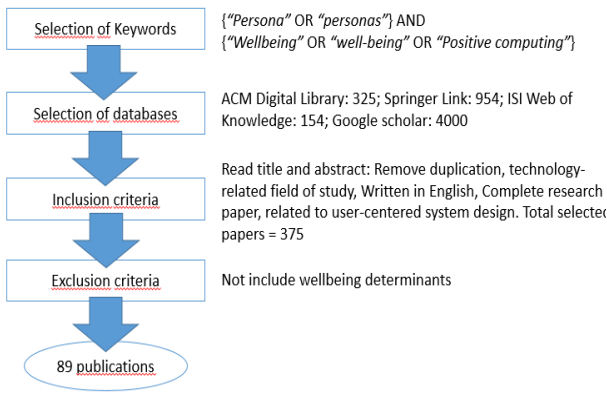

**Figure 1. Paper selection process.**

The selected articles show that the majority of publications comes from Proceedings Paper (71.9 %), then Journal Papers (18 %) as the second and third come from the category of Book Chapters (10.1 %). The use of a simple regression line shows that the improvement in the trend line both in the number of publications per year (see Figure 2) and in Figure 3, the percentage of publications related to personas per year at a non-specific Human-Computer Interaction (HCI) conferences indicates the increase in the use of personas outside the HCI community.

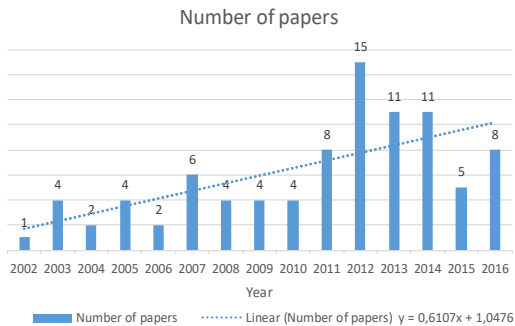

**Figure 2. Distribution of publications.**

The context of our study is the creation process of personas, and because the core activities (identifying, grouping, defining, and synthesizing) of creation are not directly related to the well-being, we then correlated the well-being with the object of each activity in the creation process. Based on the personas creation process [6], we identified the main activities that related to enhancing well-being determinants: grouped the roles, identified behavioral attributes and behavioral patterns, synthesized characteristics and defined the user goal. Next, we used positive personas [29] as the basic framework to develop wellbeing-driven design process. Finally, we elaborated the personas creation process with positive personas to propose wellbeing-driven personas intervention. Figure 3 shows an overview of the study result.

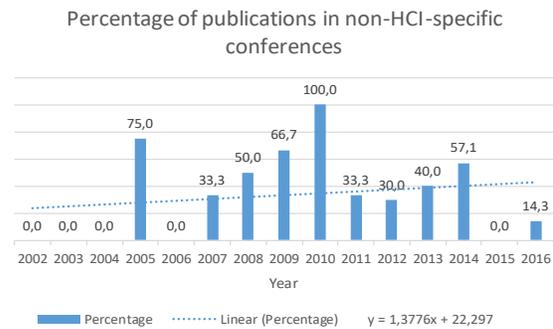

**Figure 3. Percentage of publications in non-HCI specific conferences each year.**

The wellbeing-driven creation process of personas consists of three main parts (see figure 4). The first part is the empathizing phase, where the main activity is to collect user data. The data can be gathered from an interview, observation of the real world or from available digital data (e.g., social media data, digital news, email conversation, image, and video).

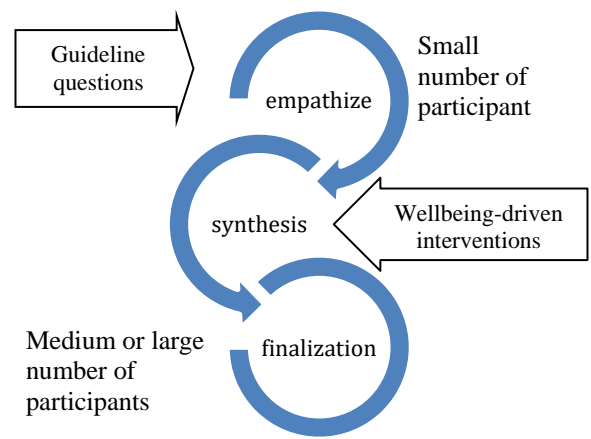

**Figure 4. The wellbeing-driven personas creation process.**

A set of questions in Table 1 will be used to guide the data collection and to minimize repeated work of data collection because of the lack of required information for the next phase. The questions can also be used to rebuild existing personas to highlight the aspect that can be correlated to wellbeing determinants. By using the guideline question, the system designers can communicate with persons who collect the data and ask the data collectors for information related to the question as well as information for primary personas related to demographical questions, expected functionality of system and positive emotional experiences (gains) as well as the barriers and difficulties (pains) for specific tasks or activities.

In the synthesis phase, all collected user information undergoes analysis and classification based on certain aspects of positive personas [29]. Furthermore, wellbeing-based interventions are proposed to support the classification process of personas including the process for identifying the wellbeing determinants, identifying the optimum duration for technological use, as well as for identification of digital and non-digital technological interaction to support learnability of the proposed system design.

In the finalization process, the proposed personas classification and the correspondence list of identified information related to targeted activity can be verified with the targeted user. Because the main aim of the wellbeing-driven system design is to promote subjective wellbeing, the Q-Methodology [3] can be used by system developers to develop personas based on the sorting process of user subjectivities on the particular issue [15, 26] to wellbeing [14]. The Q-method also provide both qualitative and quantitative result for further use in the design process [15, 26]. The list of identified information from several users can act as the Q-statements and the proposed personas type as the question for targeted personas classification [7]. Through the Q-methodology pattern of Q-statements can be mapped directly to different user types [7].

Next, we present the guideline questions and the wellbeing-driven interventions for the proposed positive-personas creation process.

## POSITIVE PERSONAS AS A TOOL FOR WELLBEING-DRIVEN DESIGN

The guideline questions for positive-personas consist of five human aspects (role, motivational orientation, goal orientation, operational orientation, and time orientation) [29] related to the object in the creation process of personas. The five aspects were chosen because the aspects describe items related to user activities that appear simple to the user but can contribute to the well-being of the user [22, 23] and can be integrated into the design of the user experience [2].

First, concerning the motivational aspect. Ryan & Deci proposed three basic physiological needs (competence, relatedness, and autonomy) as self-determination theory (SDT) [32] and explained why a specific activity that correlated to the underlying psychological motivation could improve human well-being [5, 32].

| \ | \ | \ | \ |
|---|---|---|---|
| **Motivation** | | | |
| *Elements of positive-personas* | *Basic question:* Why? | Artifact element Autonomy, competence, relatedness | Human element Motivational orientation |
| *Guideline Questions* | Why does the persona do the activity? Why do they need a digital solution for the activity and why not? | | |
| **Instrumental aspect** | | | |
| *Elements of Positive-Personas* | *Basic question:* What? | Artifact element Compassion, self-esteem | Human element Goal orientation |
| *Guideline Questions* | What is the technology-related goal of the selected Personas activity? What is the correlated determinant of well-being in the user goal that can be supported? | | |
| **Operational aspect** | | | |
| *Elements of Positive-Personas* | *Basic question:* How? | Artifact element Handling and physical aspect Adaptive aspect | Human element Learning aspect and Adaptation |
| *Guideline Questions* | How does the persona use a particular tool/technology to perform the activity (current technological usage or potential technology with which the user already has experience)? | | |
| **Timing aspect** | | | |
| *Elements of Positive-Personas* | *Basic question:* When? | Artifact element Flow (Dosage and variety) | Human element Time orientation (past, present, or future) |
| *Guideline Questions* | When the activity is performed (intensity and variation)? | | |
| **Social embedding** | | | |
| *Elements of Positive-Personas* | *Basic question:* Who? | Artifact element Social support | Human element Social Role |
| *Guideline Questions* | Who is the social support person when the user carries out the activity (or the possibility to be connected with the user in the future)? Whom does the persona want to get in contact with (regarding the context)? | | |

**Table 1. Aspects and guideline questions of wellbeing-driven personas adapted from positive personas [25].**

Not all aspect should be digitalized, there is a need for equity in the system design to support the wellbeing, and it depends on the user experience and user competence [2]. The designer should take care to provide equity of the digital interaction between the user and the system [19, 35]. Therefore, two related questions to the motivational aspect are proposed.

The second, the timing aspect is related to the flow of activity dosage and the variety of activity that can improve well-being [22, 23]. The designer needs to take attention to the flow of activity and to ensure when is the right time of activity with the given information about social supports, goals, and capability is needed to perform by the user.

The third aspect is the role of the user personas; each person has a role in social life. Concerning the social role, the well-being determinant can be facilitated by identifying the social relationship. The aim is both to increase the sense of social attachment and to convey a sense of gratitude, which is one of the best predictors of well-being [39]. The information of the third aspect leads to the question related to the social embedding aspect of the personas about social support and the potential of a further person to support the social life of the targeted personas.

Fourth, Instrumental aspect is related to the goal of the user activity. It is essential that designers collect information about what are the user goal that can be facilitated by technology or any specific type of wellbeing determinants is essential for the user to achieve the goal [29].

The last aspect is about how an activity can be facilitated by the physical and cognitive aspect that can enhance the feeling of competence [10, 31] to operate the technology and interact with the design. The complete guideline questions and the correlated aspects of positive-personas are listed in table 1. Next, we present the interventions based on positive personas into the creation process of personas.

**Positive Personas Design Interventions**

The process of integration of wellbeing determinants was starting by defining the object of each process of personas creation. The similarity of the object in positive personas than map into the creation process of personas adapted from [6]. Then, we related each object in the activity with the human aspect of positive personas. The interventions also consist of five essential activities that integrated into the creation process of personas. The creation process of personas starting from a) group interview, b) identification of behavioral variables, c) Mapping subject based on behavioral variables, d) Identification of pattern, e) Synthesize goals, f) check for redundancy and completeness, g) design personas types, h) Expand the description of personas [6].

The first intervention related to 1) interview of group subject by role. In the interview phases, the system designer needs to identify the social interaction of each role that can support the usage-behavior and improvement of the user well-being.

Intervention related to 2) identification of behavioral variables is finding specific behavioral that can improve the feeling of competence, autonomy, relatedness (CAR) [32] of the user while doing an activity related to the product. Intervention related to the 4) identification of pattern by taking attention to the frequency, timing, sequence, and a variety of usage behaviors that can improve the joyful flow of user experience. Finally, two interventions are integrated into 5) synthesize the goals and user characteristics: identification of the physical and cognitive condition of the user. As well as past technological experience and the future expected technological environment that can facilitate a high level of self-compassion and self-esteem.

| Creation process | Wellbeing interventions |
|---|---|
| Group interview subjects by role | Identify the social interaction of each role that supports the usage-behavior and improvement of the user wellbeing. |
| Identify behavioral variables | Identify behaviors that can improve the feeling of competence, autonomy, relatedness (CAR) of the user, while doing an activity related to the system/product. |
| Map interview subjects to behavioral variables: no proposed intervention | |
| Identify significant behavior patterns | Identify frequency, timing, sequence, and a variety of usage behaviors that can improve **joyful flow of usage experience**. |
| Synthesize characteristics and define goals | • Identify current **physical and cognitive** conditions that can support adaptation and the feeling of CAR while performing the targeted usage activity.<br>• Identify **past and future conditions** that can facilitate a high level of **self-compassion and self-esteem** in obtaining the usage goals |
| Check for redundancy and completeness: no proposed intervention | |
| Designate personas types: no proposed intervention | |
| Expand description of attributes and behaviors: no proposed intervention | |

**Table 2: Wellbeing-driven interventions to personas creation process adapted from [6].**

The interventions do not include specific behavioral attributes of personas. Instead of general personas attributes related to well-being, the interventions can be applied to any type, use, and creation process of personas. Table 2 shows the interventions of positive personas into the personas creation process. The positive personas extend the current focus of personas from summarizing the usage goals and behaviors [6] to focus on well-being mediators that drive those goals and behaviors. By utilizing positive personas, the designers need to relate the behavioral attributes to well-being determinants that can facilitate collaboration among the team to communicate design decision [25, 33] as thoroughly evaluate selected requirements from well-being based views [30].

## DISCUSSION

### Example of Use of Positive Personas

This section describes the use of positive personas in different scenarios - the use of scenarios [6, 20] was chosen to validate the feasibility of positive personas and thus provide initial proof of concept. Some conditions in which we recommend the use of personas are, first, well-being is used as a collective design goal. Positive personas provide a tool to guide the development team to integrate well-being into important attributes of personas in the earlier phases of the design [9]. Second, a persona is used with high intensity in cross-functional collaboration. At this point, positive personas can be used as a communication tool that drives the goal of communication and bridge the communication gap by focusing on one collective goal to promote well-being [33].

Lastly, the target user has needs related to physical and cognitive capability. Positive personas can be used to integrate demographical attributes into the design and provide a flow of experience by using the system according to the physical and cognitive capability of the user [23].

An example of use on a result of the process for using the guideline questions and the interventions is given in figure 5. To create this persona, we analyzed the situation of elderly people in times of demographic changes. From different national statistics, it is known that the proportion of elderly people in the whole society grows and that many families live spread all over the country. From several unstructured interviews with elderly people. Operators and caregivers from retirement homes we received the information that loneliness is one important fear of elderly people leading to sadness and melancholy. Randomized research in web forums, especially for elderly persons, supported these findings.

In this example, we elaborate persona "Judith Einzig" from the website[1] with further details were taken from publicly available personas that describe a different type of older adults in European countries [38]. After identifying the problems related to loneliness in older adults, we then aim to design a system that supports social communication activity, where the elderly can spend time together with other people. Based on the collected data from personas, we synthesized the information and mapping into positive personas based on guideline questions. More details on the developed system that allows grandparents to play a game with their grandchildren over distance with a tangible user interface can be found in [4].

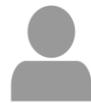

**AIRA Suzanne, 62, Female**
Activity: Spending time together, communication, familiy and friendship activities

Motivational aspect (**Why** does the persona do the activity?):
Continue to be an active part of the family (experience activity with grandchildren). Reducing sadness due to loneliness

Instrumental aspect (**What** is the aim of the technology-related activity?):
Video telephony and playing with grandchildren

Operational aspect (**How** does the persona use a particular tool/technology to perform the activity?):
As natural as possible with the previous technological environment and the interaction approach: Object of interaction and people should appear in realistic form or size. Play with real features instead of virtual ones.

Timing aspect (**When** is the activity performed? (intensity and variation)):
Approx. one hour per week and should be plannable.
(needs an exact appointment)

Social embedded (**Who** does the persona want to be in contact with (or the possibility to be connected with the user in the future)?
Family members who live far away. Senior Community. Nursing staff. Technical expert to assist with system installation.

**Figure 5. The of use guideline questions for positive personas.**

### Wellbeing-oriented Design Towards a Humane Digitization

The main purpose of personas is to help the designer develop a user interface and interaction according to the criteria of the user in a simple way [6]. Therefore, the adoption of a positive persona has implications into the overall design process. We grouped the implication of personas by following the main processes in the user interface design adapted from ISO 9241-210:2010 as presented in Table 3.

In designing the positive computing experience, Calvo & Peter mentioned that personas could be used in the technological design process to strengthen the determinant of well-being [5]. The design process in positive computing consists mainly of user studies and evaluation of existing systems that are linked closely to contextualization and research-based activity [5].

We argue that in order to humanize the design of a digital system in the age of digital transformation, fostering well-being determinants can lead to a more meaningful digital interaction experience [5]. The presence of multiple well-being determinants in positive computing design is a challenge for system designers to determine which technology-related well-being factors should be prioritized for humanizing system design. Positive personas can be used to find out which aspects have not been maximized to support the well-being determinants of positive computing.

---
[1] http://elderlypersonas.tech-experience.at  (last accessed: 16.01.2019)

| Design Process | Design implications |
|---|---|
| User Study | Direct impact to the system design: Driving the collection of user information by focusing both on usage behavior and wellbeing mediators. |
| User Modelling | Direct impact to the system design: Detailing information about the wellbeing determinants that related to personas attributes, for instance: social supports, physical and cognitive capabilities. |
| Design requirements and specifications | Indirect impact to the system design: Focusing on design requirements that brings two outcomes. Experiential outcomes related to joyful activity and instrumental outcomes related to meaningful activity. |
| Execution and evaluation | Indirect impact to the system design: Developing an evaluation method based on positive personas and subjective wellbeing. |

**Table 3. Implications of positive personas to overall design process.**

Therefore, positive personas can be employed to govern the prioritization of particular well-being determinant. For example, a) a system supports the aspect of social embedded to increase the wellbeing determinant of social connectedness (Social Media App), the aspect of timing must also be considered, e.g., by reminding of the limitation of the time of interaction with technologies to reduce technostress and tech addiction [18]. b) A system design that supports the inclusion of users with disabilities (for the operational aspect) aims to make the system usable for a variety of users, both for people with disabilities and users without operational restrictions [1].

In this study, we, therefore, aim to humanize the design of digital systems with the help of positive personas by focusing on all five aspects of positive personas (motivational, instrumental, operational, timing, and social embedded) and not just one particular aspect. A system designer can collect and analyze these five aspects to support the determinants of well-being and develop a more meaningful system for human life.

**OUTLOOK AND RECOMMENDATIONS**

This paper can contribute to theoretical and practical applications of design techniques and methods of a system by proposing a wellbeing-based creation process of personas. Our approach utilized the established positive personas, and we present guideline questions related to the aspects of social supports, intrinsic motivations, and the flow of activities, the physical and cognitive conditions of the user.

Furthermore, we provide intervention activities based on the positive personas model to integrate well-being elements into the creation process of personas. The positive persona concept was helpfully used in research projects of the authors, but general validity and benefits need to be proven by further case studies and projects. The main concept and the proposed interventions are flexible and can easily be adapted to other contexts. A deeper understanding of opportunities and limitations in the practical usage and project business will be part of our future research.

**ACKNOWLEDGMENTS**

The research project is funded by the MIWF (Ministry for Innovation, Science, and Research of the State of NRW) as part of a project for the positive computing institute of the Hochschule Ruhr West University of Applied Sciences.